\documentclass[american,british]{article}
\usepackage[T1]{fontenc}
\usepackage[utf8]{inputenc}
\usepackage{geometry}
\usepackage{color}
\usepackage{array}
\usepackage{booktabs}
\usepackage{amsmath}
\usepackage{amssymb}
\usepackage{graphicx}
\usepackage{setspace}
\usepackage[authoryear]{natbib}
\makeatletter

\providecommand{\tabularnewline}{\\}

\@ifundefined{date}{}{\date{}}
\usepackage[page,header]{appendix}
\usepackage{titletoc}
\usepackage{lipsum}
\usepackage{chngcntr}

\usepackage{amsfonts}

\let\oldmarginpar\marginpar
\renewcommand\marginpar[1]{\-\oldmarginpar[\raggedleft\tiny #1]%
{\raggedright\footnotesize #1}}

\usepackage{array}
\newcolumntype{C}[1]{>{\centering\let\newline\\\arraybackslash}m{#1}}

\usepackage{hyperref}
  \hypersetup{colorlinks=true,citecolor=blue}

\usepackage{changepage}

\renewcommand{\marginpar}[1]{}

\renewcommand\section{\@startsection {section}{1}{\z@}%
                                   {-3.5ex \@plus -1ex \@minus -.2ex}%
                                   {2.3ex \@plus.2ex}%
                                   {\normalfont\LARGE\bfseries}}% from \Large
\renewcommand\subsection{\@startsection{subsection}{2}{\z@}%
                                     {-3.25ex\@plus -1ex \@minus -.2ex}%
                                     {1.5ex \@plus .2ex}%
                                     {\normalfont\Large\bfseries}}% from \large
\renewcommand\subsubsection{\@startsection{subsubsection}{3}{\z@}%
                                     {-3.25ex\@plus -1ex \@minus -.2ex}%
                                     {1.5ex \@plus .2ex}%
                                     {\normalfont\large\bfseries}}% from \normalsize
\allowdisplaybreaks

\@ifundefined{showcaptionsetup}{}{%
 \PassOptionsToPackage{caption=false}{subfig}}
\usepackage{subfig}
\makeatother

\usepackage{babel}
\begin{document}
\title{Multigenerational Inequality\thanks{This paper has previously been published in the \emph{Research Handbook
on Intergenerational Inequality} edited by Elina Kilpi-Jakonen, Jo
Blanden, Jani Erola, and Lindsey Macmillan, Edward Elgar Publishing
Ltd, https://doi.org/10.4337/9781800888265.00015. It is deposited
under the terms of the CC BY-NC-ND license. I am grateful for helpful
comments and suggestions from Adrian Adermon, Jo Blanden, Anders Björklund,
Lola Collado, Maximilian Longmuir, Martin Nybom and Salvatore di Salvo.
Financial support from MICIU/AEI (RYC2019-027614-I, CEX2021-001181-M)
is gratefully acknowledged. }}
\author{Jan Stuhler\\
Universidad Carlos III de Madrid}
\maketitle
\begin{abstract}
\begin{spacing}{1.2}
\thispagestyle{empty}A growing literature provides evidence on \emph{multi}generational
inequality -- the extent to which socio-economic advantages persist
across three or more generations. This chapter reviews its main findings
and implications. Most studies find that inequality is more persistent
than a naive iteration of conventional parent-child correlations would
suggest. We discuss potential interpretations of this new ``fact''
related to (i) latent, (ii) non-Markovian or (iii) non-linear transmission
processes, empirical strategies to discriminate between them, and
the link between multigenerational and assortative associations.
\end{spacing}

\begin{singlespace}
\bigskip{}

\hspace{-0.5cm}Keywords: Multigenerational inequality, assortative
mating, distant kins\bigskip{}
\bigskip{}
\bigskip{}
\bigskip{}
\end{singlespace}
\end{abstract}
\setcounter{page}{1}

\section{Introduction}

Studies of social mobility often focus on two generations, measuring
how one\textquoteright s education, income or other outcomes are associated
with that of one\textquoteright s parents. But how persistent are
socio-economic inequalities in the long run, across multiple generations?
A naive extrapolation from conventional parent-child estimates would
suggest that long-run persistence is low -- that the influence of
family background declines geometrically and washes out over three
or four generations. However, a growing literature provides direct
evidence on \emph{multigenerational inequality}. This chapter reviews
this evidence and its implications. 

We first illustrate why multigenerational regression coefficients
deviate from the iterated product of the corresponding parent-child
coefficients (Section \ref{sec:The-iterated-regression}), even though
the ``iteration'' of coefficients may appear natural in a regression
framework. Indeed, this \emph{iterated regression fallacy} has been
a common source of misinterpretations in intergenerational research
and other contexts. The reason why parent-child correlations may not
be very informative about multigenerational persistence is that they
measure only a descriptive rather than a structural relationship. 

We then review recent empirical evidence on multigenerational inequality
(Section \ref{sec:Multigenerational-inequality}). A robust pattern
across studies is that inequality is more persistent than a naive
iteration of the parent-child correlations would suggest. Put differently,
the coefficient on grandparents in a child-parent-grandparent regression
tends to be positive: grandparent status predicts child status, even
conditional on parent status. However, this \textquotedblleft excess
persistence\textquotedblright{} partly reflects the omission of important
characteristics of the parent generation; a particularly vital omission
is the second parent.

Less clear is how this new empirical ``fact'' should be interpreted.
We highlight three potential interpretations related to (i) latent,
(ii) non-Markovian and (iii) non-linear transmission processes (Section
\ref{sec:Interpretations}). Some parental influences are inherently
unobservable, and such latent processes could generate multigenerational
persistence. Alternatively, perhaps ``grandparents matter'' in a
causal sense, exerting an influence on their grandchildren that is
distinct from parental influences. As a third possibility, transmission
processes could be non-linear or vary across families. While non-linearities
have been studied in other contexts, its multigenerational implications
have received less consideration.

We also link this evidence to earlier work on sibling correlations,
which suggests that parental characteristics such as schooling or
income account only for a minor part of the family and community influences
that siblings share (\citealp{BjorklundSalvanes2011HB}; \citealp{JenkinsJaenttHandbook201x}).
Intuitively, as family background cannot be captured by one single
variable, intergenerational associations may reflect only the \textquotedblleft the
tip of the iceberg\textquotedblright{} of family background effects
(\citealp{BjorklundJaentti2012aa}). We argue that multigenerational
associations reflect this same insight, making those associations
meaningful even if their absolute size remains limited. 

One intriguing implication of strong multigenerational associations
is that assortative mating must be strong, too (Section \ref{sec:Assortative-Matching}).
Conventional measures of assortative mating imply rapid regression
to the mean across generations, which would be at odds with high multigenerational
correlations. Consequently, multigenerational studies not only shed
light on the long-term persistence of inequality, but also offer novel
insights into fundamental aspects of \emph{inter}generational transmission,
like the level of sorting within a population.

For brevity, this chapter omits many important questions. The stylised
models we consider are ``mechanical'', abstracting from economic
choices and behaviours. We do not address policy or normative questions,
such as whether multigenerational correlations are ``too high''.
Instead, this chapter focuses on basic empirical facts and their potential
interpretations, which might inform future work. It is complementary
to an earlier review by \citet{AndersonSheppardMonden2017}. While
they provide a systematic summary of multigenerational estimates,
this chapter focuses on the interpretation of such estimates and relation
to other measures of social mobility. Other insightful discussions
of multigenerational mobility include \citet{Pfeffer2014}, \citet{Solon_2018}
and \citet{Breen2018}, and some sections of this chapter draw on
\citet{Stuhler2012} and \citet{HHBlandenDoepkeStuhler23}.

\section{The Iterated Regression Fallacy\label{sec:The-iterated-regression}}

Our understanding of intergenerational processes has been shaped by
theoretical and empirical research involving just two generations,
parents and children (\citealt{Mare2011}). It is therefore instructive
to first consider why an extrapolation from the available parent-child
evidence may not be very informative about the persistence of socio-economic
status across multiple generations (in the ``long run'').

The degree of status persistence between \textit{parents} and their
children is often measured by the slope coefficient in a linear regression
of outcome $y$ in offspring generation $t$ of family $i$ on the
parental outcome in generation $t-1$,
\begin{equation}
y_{it}=\alpha+\beta_{-1}y_{it-1}+\varepsilon_{it}.\label{eq:IGE}
\end{equation}
For example, if $y$ is the logarithm of income then $\beta_{-1}$
captures the \textit{intergenerational elasticity}\emph{ of income};
a high elasticity represents low mobility. For simplicity we assume
below that $\beta_{-1}$ remains constant across generations, but
the arguments extend to non-stationary environments.

How does the coefficient from this Galtonian regression across two
generations compare with the coefficient across three or more generations?
The idea that the latter equals the square of the former, so that
persistence declines geometrically, may appear as a natural consequence
of regression: if $\beta_{-1}$ captures to what degree deviations
from the mean tend to be passed from parents to children then we might
expect $(\beta_{-1})^{2}$ to represent their expected extent after
being passed twice from parents to children, between \textit{\emph{grandparents}}
and their grandchildren. Formally, we may use equation (\ref{eq:IGE})
to rewrite the grandparent-grandchild elasticity $\beta_{-2}$ as
\begin{align}
\beta_{-2} & \equiv\frac{Cov(y_{it},y_{it-2})}{Var(y_{it-2})}=\frac{Cov(\beta_{-1}y_{it-1}+\varepsilon_{it},y_{it-2})}{Var(y_{it-2})}\overset{?}{=}(\beta_{-1})^{2}.\label{eq:IGE_iter}
\end{align}
The fallacy is in the last step: while $\varepsilon_{it}$ is by construction
uncorrelated to $y_{it-1}$, it is not necessarily uncorrelated with
grandparental status $y_{it-2}$. Intuitively, the coefficient $\beta_{-1}$
in equation (\ref{eq:IGE}) captures only a statistical, not a structural
association.\footnote{Importantly, equation (\ref{eq:IGE_iter}) may fail to hold even in
a Markovian world, in which outcomes depend only on the previous generation
\citep{Mare2011}. We illustrate this argument in Section \ref{sec:Interpretations}. }

This ``\textit{iterated regression fallacy}'' (\citealp{Stuhler2012}),
i.e. the belief that regression toward the mean between two periods
implies iterated regression across multiple periods, is a common misconception.
\citet{Bulmer2003} describes how Francis Galton fell fault of it
in his influential work on linear regression and \citet{Nesselroade1980}
discuss its prevalence in psychological research (using the label
\textquotedblright expectation fallacy\textquotedblright ). As discussed
in the next section, a naive iteration of parent-child correlations
tends instead to understate the extent of multigenerational inequality.

\section{Multigenerational Inequality\label{sec:Multigenerational-inequality}}

How persistent are socio-economic inequalities? A string of recent
studies provide a new perspective on this question by tracking families
over multiple generations. Spurred by the increased availability of
suitable data, research on multigenerational mobility has surged nearly
simultaneously in economics (e.g., \citealp{LindahlPalme2014_IGE4Generations}),
sociology (\citealt{Tak-Wing-Chan:2013aa}), demography (\citealt{Mare2011}),
and economic history (\citealt{Dribe:2016aa}). However, different
studies emphasise different interpretations, a point to which we return
in the next section. \citet{AndersonSheppardMonden2017} provide a
systematic review of earlier studies, and is complementary to the
more selective presentation here.

\subsection{Measuring multigenerational inequality\label{sec:Measuring}}

Multigenerational evidence tends to be presented in one of two distinct
forms. We may compare the relative size of inter- and multigenerational
correlations, i.e. whether multigenerational correlations are larger
or smaller than the naive iteration of parent-child correlations would
suggest,
\begin{equation}
\beta_{-k}\lesseqgtr\left(\beta_{-1}\right)^{k}\label{eq:inter_vs_multi}
\end{equation}
where $\beta_{-k}$ is the multigenerational correlation between generation
$t$ and generation $t-k$, for $k>1$. Alternatively we may estimate
a multivariate regression of the form
\begin{equation}
y_{it}=\alpha+\beta_{p}y_{it-1}+\beta_{gp}y_{it-2}+...+\varepsilon_{it}.\label{eq:3-generation-reg}
\end{equation}
and study the sign and magnitude of the slope coefficients on grandparents
or earlier ancestors.\footnote{One interesting observation is that the addition of grandparents or
other ancestors often contributes little in an $R^{2}$ sense, even
if the corresponding slope coefficients are large. We return to this
observation in Section \ref{sec:Caveats-and-Criticisms}.} The distinction is just a presentational one: using the Frisch-Waugh-Lovell
theorem, the coefficient $\beta_{gp}$ in the three-generation regression
(\ref{eq:3-generation-reg}) can be re-expressed as (see \citealp{ECOJ:ECOJ12453})
\begin{equation}
\beta_{gp}=\frac{\beta_{-2}-(\beta_{-1})^{2}}{1-(\beta_{-1})^{2}},\label{eq:Duality}
\end{equation}
such that we have ``excess persistence'' in the sense of $\beta_{gp}>0$
if and only if $\beta_{-2}>\left(\beta_{-1}\right)^{2}$. Given this
duality, there is a close link between studies providing bivariate
estimates, as in (\ref{eq:inter_vs_multi}), and those focusing on
multivariate estimates, as in (\ref{eq:3-generation-reg}). 

\subsection{Multigenerational evidence\label{sec:Multigenerational-evidence}}

\begin{table}
\caption{Selected Multigenerational Studies}
\label{table:studies}
\begin{centering}
\begin{tabular*}{1\textwidth}{@{\extracolsep{\fill}}>{\raggedright}p{0.16\textwidth}>{\raggedright}p{0.12\textwidth}>{\raggedright}p{0.12\textwidth}>{\raggedright}p{0.19\textwidth}>{\raggedright}p{0.36\textwidth}}
\toprule 
{\small{}Study} & {\small{}Sample} & {\small{}Main outcomes} & {\small{}Excess persistence} & {\small{}Remarks}\tabularnewline
\midrule
{\small{}\citet{WarrenHauser1997}} & {\small{}US (Wisconsin)} & {\small{}Education, Occupation} & {\small{}No, cond. on parent characteristics} & {\small{}Conditioning tests}\tabularnewline\addlinespace
{\small{}\citet{LindahlPalme2014_IGE4Generations}} & {\small{}Sweden (Malmö)} & {\small{}Education, Income} & {\small{}Yes, $\beta_{-k}>\beta^{k}$ and $\beta_{gp}>0$} & {\small{}Link up to four generations for education}\tabularnewline\addlinespace
{\small{}\citet{AdermonLindahlWaldenstroem2018}} & {\small{}Sweden} & {\small{}Wealth} & {\small{}Yes,  $\beta_{gp}>0$} & {\small{}Study bequests and wealth, earnings, education up to four
generations}\tabularnewline\addlinespace
{\small{}\citet{ECOJ:ECOJ12453}} & {\small{}Germany } & {\small{}Education, Occupation} & {\small{}Yes, $\beta_{-k}>\beta^{k}$ and $\beta_{gp}>0$} & {\small{}Latent factor model, conditioning tests, education up to
four generations}\tabularnewline\addlinespace
{\small{}\citet{PfefferKillewald2018}} & {\small{}US} & {\small{}Wealth} & {\small{}Yes, $\beta_{-k}>\beta^{k}$ and $\beta_{gp}>0$} & {\small{}Study role of bequests, educational attainment, marriage,
homeownership, and business ownership}\tabularnewline\addlinespace
{\small{}\citet{SheppardMonden2018}} & {\small{}21 countries (SHARE)} & {\small{}Education} & {\small{}Yes, $\beta_{gp}>0$} & {\small{}Test for grandparent, contact and interaction effects}\tabularnewline\addlinespace
{\small{}\citet{NeidhoeferStockhausen}} & {\small{}US, UK and Germany} & {\small{}Education} & {\small{}Yes, $\beta_{-k}>\beta^{k}$ and $\beta_{gp}>0$ (UK and
Germany)} & {\small{}Conditioning tests, multigenerational trends, latent factor
model}\tabularnewline\addlinespace
{\small{}\citet{colagrossi2020like} } & {\small{}28 EU countries } & {\small{}Education, Occupation} & {\small{}Yes, $\beta_{-k}>\beta^{k}$ and $\beta_{gp}>0$ (most countries)} & {\small{}Conditioning tests, latent factor model}\tabularnewline\addlinespace
{\small{}\citet{Engzell:2020aa}} & {\small{}Sweden} & {\small{}Income} & {\small{}Unconditional $\beta_{gp}>0$, conditional $\beta_{gp}\approx0$ } & {\small{}Conditioning tests, heterogeneity of multigenerational associations}\tabularnewline\addlinespace
{\small{}\citet{hallsten2020shadow}} & {\small{}Sweden (Skellefteå and Umeå)} & {\small{}Education, Occupation, Wealth} & {\small{}Yes, $\beta_{-k}>\beta^{k}$ and $\beta_{gp}>0$} & {\small{}Up to seven generations and $5^{th}$-order cousins}\tabularnewline\addlinespace
{\small{}\citet{modalsli2021multigenerational}} & {\small{}Norway} & {\small{}Occupation, Income} & {\small{}Yes, $\beta_{-k}>\beta^{k}$ and $\beta_{gp}>0$} & {\small{}Study contact effects, multigenerational trends}\tabularnewline\addlinespace
{\small{}\citet{li2023multi}} & {\small{}China} & Education & {\small{}Yes, $\beta_{-k}>\beta^{k}$ and $\beta_{gp}>0$} & {\small{}Conditioning tests, contact effects, multigenerational trends }\tabularnewline\addlinespace
\bottomrule
\end{tabular*}
\par\end{centering}
{\footnotesize{}\smallskip{}
Notes: Selected studies with multigenerational family links. ``Conditioning
tests'' correspond to estimates of $\beta_{gp}$ in equation (\ref{eq:3-generation-reg})
with different sets of parental controls.}{\footnotesize\par}
\end{table}
Until recently, little multigenerational evidence has been available.
\citet{Hodge1966} warns that mobility may not be well described by
a first-order Markov process in which child outcomes depend only on
the parent generation.  Studying families over three generations
in Wisconsin, \citet{WarrenHauser1997} show that the occupational
status of grandparents is not very predictive of their grandchildren's
status once father's education, occupation and earnings, and mother's
education, are controlled for.  However, their estimates are not
very precise due to the limited size of their sample. Similarly, \citet{ErolaMoisio2007}
report that conditional on parents' \emph{class} (see also \citealt{heath2024intergenerational},
published in the same volume as this chapter), the grandchildren's
social class is nearly independent from the grandparents' class in
Finland, while \citet{Tak-Wing-Chan:2013aa} find a more robust conditional
association in British data. See \citet{HertelSamberg2013} for further
discussion of these studies.

\citet{LindahlPalme2014_IGE4Generations} combine survey data from
the Swedish ``Malmö study'' with administrative data to track earnings
for three generations and educational attainment over four generations.
They find that multigenerational persistence is much higher than would
be predicted from the iteration of regression estimates for two generations,
i.e. $\beta_{-k}>\left(\beta_{-1}\right)^{k}$ and $\beta_{gp}>0$
in the three-generation regression (\ref{eq:3-generation-reg}). The
size of the coefficient on grandparents' (standardised) earnings is
about one quarter of the corresponding coefficient on father's earnings,
which is also the median ratio of $\beta_{gp}/\beta_{p}$ across 40
research articles reviewed by \citet{AndersonSheppardMonden2017}.
The results by Lindahl et al. received much attention. Although restricted
to one Swedish region, their data are of high quality and contain
earnings. Moreover, their findings are at odds with a well-known prediction
by \citet{becker1986human} that\foreignlanguage{american}{ ``\emph{Almost
all earnings advantages and disadvantages of ancestors are wiped out
in three generations}''.} 

\citet{ECOJ:ECOJ12453} interpret multigenerational correlations in
educational and occupational status in different German samples through
the lens of a latent factor model (see Section \ref{subsec:Latent-transmission-processes}).
The implied parent--child correlation in ``latent'' advantages
is about 0.6, nearly 50 percent larger than the parent-child correlation
in years of schooling in their samples. Applying a similar approach
on harmonised survey data, \citet{NeidhoeferStockhausen} find slightly
higher latent persistence around 0.7 for Germany and slightly lower
persistence rates for the US and UK, while \citet{colagrossi2020like}
estimate a mean rate of latent persistence of 0.66 in standardised
educational outcomes across 28 European countries. 

Similar patterns are found for other outcomes, such as wealth (see
also Chapter 7 by Lersch, Longmuir and Schnitzlein). Using Swedish
data, \citet{AdermonLindahlWaldenstroem2018} show that grandparents'
wealth is predictive of grandchildren\textquoteright s wealth, above
and beyond parent wealth. This pattern is even more pronounced in
studies by \citet{Boserupetal2013} for Denmark and \citet{PfefferKillewald2018}
for the US. While it could be partially explained by direct bequests
from grand- to grandchildren that ``skip a generation'' (\citealp{Mare2011}),
advantages associated with family wealth arise at an earlier age than
such direct bequests would imply. Conditional on parental wealth,
grandparents' wealth also predicts other outcomes of their grandchildren,
such as education and home ownership (\citealp{PfefferKillewald2018})
or school grades (\citealp{hallsten2017grand}). 

Some recent studies are able to link more than ``just'' three generations.
In particular, \citealp{hallsten2020shadow}, link administrative
data and parish records from Northern Sweden to track up to seven
generations. The observation of such long data coverage would also
allow researchers to estimate \emph{trends} in multigenerational persistence.
For example, \citet{modalsli2021multigenerational} links up to five
generations of data in Norway, and finds substantial differences in
the strength of multigenerational persistence over time. As yet there
exists little evidence on multigenerational correlations in developing
countries, with \citet{razzu2020three}, \citet{kundu2021multigenerational}
and \citet{Celhaygallegos2025} as recent exceptions. 

Most studies find that multigenerational correlations are larger than
a naive iteration of parent-child estimates would suggest. This pattern
appears robust across countries and different socio-economic outcomes.
However, a substantial share of this ``excess persistence'' can
be explained by the omission of the second parent, and studies that
control for both maternal and paternal characteristics (e.g., \citealp{Engzell:2020aa})
tend to find a much smaller and sometimes insignificant coefficient
$\beta_{gp}$ in the child-parent-grandparent regression (\ref{eq:3-generation-reg}).
There exists only limited evidence on whether multigenerational patterns
vary across countries or groups.\footnote{Cross-country comparisons by \citet{global2018global}, \citet{NeidhoeferStockhausen},
\citet{colagrossi2020like} or \citet{Celhaygallegos2025} do find
such variation. Open questions for future research include whether
cross-country (\citealp{Blanden2011Survey}) or regional differences
(see Chapter 15 by Marie Connolly and Catherine Haeck) in parent-child
correlations are also reliable indicators of differences in multigenerational
inequality.} Most importantly, there is no consensus yet on how those patterns
should be interpreted: while some studies emphasise the role of ``latent''
transmission channels, others emphasise the causal influence of grandparents
or the extended family. We discuss potential interpretations in Section
\ref{sec:Interpretations}.

One common issue in multigenerational studies is that the marginal
distributions tend to be very different for distant ancestors. Educational
attainment is often low, income rarely observed, and occupational
classifications can be problematic if the share of farmers is high
in older generations. Moreover, very different mechanisms can generate
similar ``vertical'' transmission patterns (\citealp{cavalli1981cultural}),
making it difficult to distinguish between competing models. One alternative
explored in recent studies is to consider distant relatives in the
``horizontal'' dimension. \citet{Adermonetal2016} measure \textquotedblleft dynastic
human capital\textquotedblright{} based on a broad set of kinships
in the parent generation, including uncles and aunts, and show that
it has a much stronger association with child education than conventional
parental measures. \citet{ColladoOrtunoStuhlerKinship} show that
a single transmission model with strong assortative mating (see Section
\ref{sec:Assortative-Matching}) can fit both vertical and horizontal
kinship correlations.

Finally, some studies provide \emph{causal} evidence on multigenerational
spillovers. For example, \citet{butikofer2022breaking} show that
economic shocks affect social mobility not only in directly affected
generations, but also have indirect effects on mobility in the third
generation. Using a simple theoretical model, \citet{NybomStuhlerTrends2014}
show that even a single structural change may trigger transitional
dynamics in mobility over several generations, which can be non-monotonic.

While this chapter focuses on studies that use \emph{direct} family
links across generations, it is also related to recent \emph{name}-based
evidence by \citet{Clark2014book}, \citet{BaroneMocetti2016}, \citet{Belloc2023uy}
and others. Names are informative about multigenerational persistence
for two distinct reasons. The more obvious one is that using names,
we can link very distant generations, at least in a probabilistic
sense. For example, \citet{BaroneMocetti2016} find that the average
status of surnames correlates across five centuries, which suggests
that some components of the transmission process must exhibit high
persistence. The second, more subtle reason is that the regression
of surname averages between two generations might tell us something
about the intergenerational transmission process that is not visible
from individual-level regressions (\citealt{Clark2014book}, \citealt{ClarkCummins2014EJ}).
We return to this observation in the next section.

\section{Interpretations and Mechanisms\label{sec:Interpretations}}

How should this multigenerational evidence be interpreted? While
informative about the long-run persistence of socio-economic inequalities,
it is not directly informative about the underlying mechanisms, and
different theoretical models could generate similar multigenerational
patterns. This section reviews three potential interpretations related
to (i) latent, (ii) non-Markovian and (iii) non-linear transmission
processes.\footnote{The presentation in this section draws and extends on \citet{Stuhler2012}.
See also \citet{Lundberg2020}, who shows that different theoretical
processes could explain the relative size of sibling and cousin correlations.
We focus on purely ``mechanical'' transmission models, which can
however be viewed as the reduced form of behavioural models (see \citealp{Goldberger1989},
and \citealp{lindahl2024intergenerational}). The interpretations
reviewed here are not exhaustive; for example, \citet{Zylberberg2013}
considers a model in which dynasties move across careers, while \citet{HertelSamberg2013}
discuss the role of racial or ethnic segregation. } 

\subsection{Latent transmission processes\label{subsec:Latent-transmission-processes}}

So\textcolor{black}{me of the advantages that parents transmit to
their children may be inherently unobservable, as has long been recognised
in both the social sciences (e.g., \citealt{Duncan1969}, \citealt{Goldberger1972PathAnalysis},
\citealp{becker1979equilibrium}) and population genetics (\citealt{RiceCloningerReich1978},
\citealt{cavalli1981cultural}). }\citet{Clark2014book} and \citet{Stuhler2012}
show that \textcolor{black}{such ``latent'' transmission has also
interesting implications for the pattern of multigenerational transmission} 

To understand the basic argument, consider a simplified one-parent
one-offspring family structure, in which the transmission in generation
$t$ of family $i$ is governed by 
\begin{align}
y_{it} & =\rho e_{it}+u_{it}\label{eq:1_level_y}\\
e_{it} & =\lambda e_{it-1}+v_{it},\label{eq:1_level_e}
\end{align}
in which an observed outcome $y$ depends on latent endowments $e$
(according to \textit{returns} $\rho$), which are partially transmitted
within families (according to \textit{transferability} $\lambda$),
and where $u$ and $v$ are white-noise error terms representing market
and endowment luck, uncorrelated with each other and past values.
To simplify the presentation we drop the $i$ subscript and assume
that $e$ and $y$ are standardised with mean zero and variance one,
such that the slopes $\rho$ and $\lambda$ can be interpreted as
correlations. To make matters concrete assume that our outcome of
interest is (log) income, and that $e$ measures an individual's human
capital, although the argument can be applied in other contexts. The
parameter $\rho$ then measures the fraction of income that is explained
by an individual's own human capital, as opposed to factors or events
outside of individual control, such as market luck or market-level
shocks; and $\rho=1$ would imply that income differences are fully
explained by an individuals' own characteristics. 

Given this model, the intergenerational elasticity of income equals
\begin{align}
\beta_{-1} & =Cov(y_{t},y_{t-1})\nonumber \\
 & =\rho^{2}\lambda,\label{eq:2_layers_betaIGE}
\end{align}
and the elasticity across three generations equals
\begin{align}
\beta_{-2} & =Cov(y_{t},y_{t-2})\nonumber \\
 & =\rho^{2}\lambda^{2}.\label{eq:2_layers_betaGPIGE}
\end{align}
The extrapolation error from the iteration of the parent-child elasticity
equals
\begin{align}
\Delta & =(\beta_{-1})^{2}-\beta_{-2}\nonumber \\
 & =\left(\rho^{2}-1\right)\rho^{2}\lambda^{2},\label{eq:2_levels_extrperror}
\end{align}
which is negative if $0<\rho<1$, that is as long as income is not
perfectly determined by human capital. 

\begin{figure}[t]
\caption{Multigenerational correlations in different models}

\label{Figure_simulations}

\subfloat[A latent factor model]{
\centering{}\includegraphics[scale=0.75]{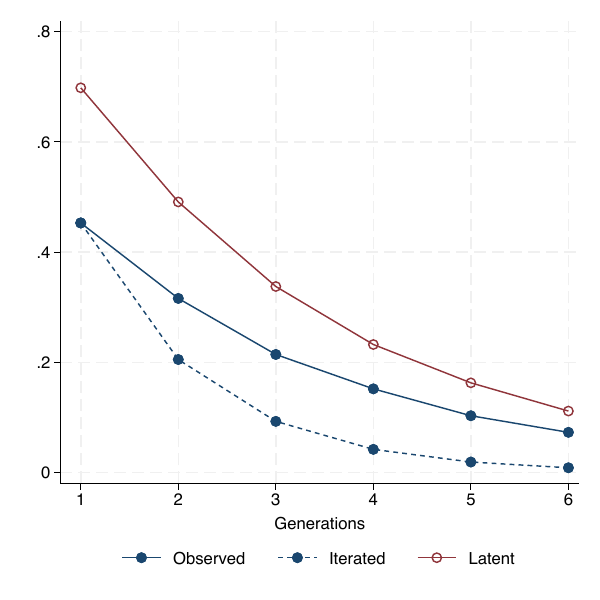}}\subfloat[A model with multiplicity]{
\centering{}\includegraphics[scale=0.75]{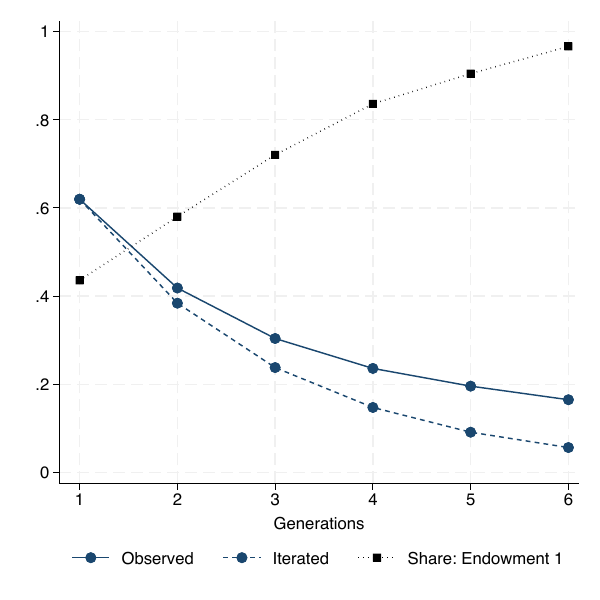}}

{\footnotesize{}\vspace{-0.5cm}
Notes: Simulated data with n=10,000 observations. Panel (a) corresponds
to the latent factor model with returns $\rho=0.8$ and transferability
$\lambda=0.7$. The solid blue (red) line corresponds to the implied
correlation in the observed outcome $y$ (latent endowment $e$).
The dashed line represents the }\emph{\footnotesize{}predicted}{\footnotesize{}
correlation based on the iteration of the parent-child correlation.
Panel (b) corresponds to the multiplicity model with two endowments
and $\rho_{1}^{2}=0.3$, $\rho_{2}^{2}=0.7$, $\lambda_{1}=0.9$ and
$\lambda_{2}=0.5$. The black dashed line corresponds to the share
of the correlation in $y$ that is explained by the first endowment.}{\footnotesize\par}
\end{figure}

Figure \ref{Figure_simulations}a provides a numerical example from
this model, assuming $\rho=0.8$ and $\lambda=0.7$, and implying
an intergenerational correlation of $\beta_{-1}=\rho^{2}\lambda\approx0.45$.\footnote{The distributions $u$ and $v$ are chosen such that {\footnotesize{}$e$
and $y$ are normally distributed with mean zero and variance one.}} A naive iteration of this parent-child correlation across multiple
generations would imply rapid regression to the mean (dashed line);
after only three generations, the iterated correlation falls below
0.1, such that the distribution of income among ancestors explains
less than one percent of the variance in their descendants' income.
But the \emph{actual} correlation in income in this model decays much
more slowly (solid blue line), falling below 0.1 after only six generations.
The reason is the strong transmission of human capital (red line),
which serves as the actual state variable in this model.

The key idea underlying this ``latent factor model'' is that the
true transmission mechanisms are distinct from the status $y$ observed
by the researcher.\footnote{The extrapolation error $\Delta$ will be particularly large when
$\rho$ is small, i.e. when the observable outcome $y$ is not a good
proxy for the latent endowments $e$.  The gap between inter- and
multigenerational correlations should therefore be particularly large
when considering outcomes that are hard to measure, such as non-cognitive
skills (e.g., \citealp{anger2017cognitive}, and \citealp{kroger2024non}).} However, the representation of this idea in equations (\ref{eq:1_level_y})
and (\ref{eq:1_level_e}) is sufficiently generic to nest several
distinct interpretations, with different implications. One possible
interpretation is that $y$ corresponds in fact to the ``true''
socio-economic status of an individual, but that status is transmitted
not directly but indirectly via other pathways. For example, income
$y$ may be a good proxy for status, but parents might transmit human
capital $e$ rather than income to their children. In this interpretation,
$\beta_{-1}$ is in fact a truthful measure of status persistence
between one generation and the next -- it is just not very informative
about persistence in the long run.

Alternatively, we may assume that $y$ is only a coarse proxy of socio-economic
status, while $e$ is the ``true'' or ``generalised socio-economic
status'' of a person (as in \citealp{Clark2014book}). For example,
$y$ may be short-run income, while $e$ may represent a broader measure
of socio-economic success. In this interpretation, $\beta_{-1}$ is
not only an inappropriate measure of multigenerational persistence;
it is not even a good measure of the \emph{inter}generational persistence
of status differences from parents to their children (which would
instead be captured by the red line in Figure \ref{Figure_simulations}a).
Measurement error in the outcome $y$ would be one important special
case  (see \citealt{Solon201413}, \citealt{ferrie2020grandparents}
and \citealp{nybom2024intergenerational}, published in the same volume
as this chapter).

The basic proposition underlying the latent factor model is intuitive,
and also consistent with earlier insights from the literature. In
particular, it is consistent with the argument that sibling correlations
are a more comprehensive measure of family background effects than
intergenerational correlations, as they capture the influence of all
the advantages that siblings share, not only the advantages encapsulated
by parental income or education (\citealp{BjorklundSalvanes2011HB}).
More generally, it is consistent with the argument that intergenerational
correlations just measure the ``tip of the iceberg'' of family background
effects (\citealp{BjorklundJaentti2012aa}). However, not all models
with latent transmission mechanisms generate high multigenerational
persistence. In particular, the Becker-Tomes model in which child
outcomes $y$ also depend on parental income can produce either high
or low multigenerational persistence (in the sense of $\beta_{-k}\lesseqgtr$$(\beta_{-1})^{k}$),
depending on parameter values.\footnote{The model will tend to produce low multigenerational correlations
(i.e., $\Delta<0$) if parental income has a strong direct effect
on child outcomes (\citealp{Stuhler2012}); moreover, $\Delta<0$
holds with certainty in simplified versions of the Becker-Tomes model
that abstract from the stochastic nature of the relation between income
$y$ and human capital $e$ (\citealp{Solon201413}).} 

\subsection{Non-Markovian and extended-family processes\label{subsec:Non-markovian-processes-and}}

A second possibility is that intergenerational transmission deviates
from a Markovian process, in that grandparents or other ancestors
have an independent influence over and above the influence of parents.
This possibility has already been considered by \citet{WarrenHauser1997},
but the role of the wider family has received renewed attention after
\citet{Mare2011} published his eloquent criticism of the ``two-generation
paradigm'' in intergenerational research (e.g., \citealt{Tak-Wing-Chan:2013aa},
\citealt{HertelSamberg2013}, \citealt{ferrie2020grandparents}). 

For illustration, assume that offspring human capital depends on both
parents and grandparents,
\begin{align}
y_{t} & =\gamma_{p}y_{t-1}+\gamma_{gp}y_{t-2}+v_{t},\label{eq:grandparent_effects}
\end{align}
where $v_{t}$ is a white-noise error term assumed to be uncorrelated
to the outcomes of parents or earlier ancestors. Note that compared
to the descriptive associations captured by equation (\ref{eq:3-generation-reg}),
we chose different symbols for the slope coefficients to indicate
that equation (\ref{eq:grandparent_effects}) has a \emph{structural}
interpretation. Of course, if this ``grandparent-effects'' model
is indeed the right model then the coefficients would coincide ($\beta_{p}=\gamma_{p}$
and $\beta_{gp}=\gamma_{gp}$), and we would observe ``excess persistence''
($\Delta<0$) iff $\gamma_{gp}>0$. 

Why might grandparents have an independent influence on their grandchildren?
\citet{Mare2011}, \citet{Tak-Wing-Chan:2013aa}, \citet{HertelSamberg2013}
and \citet{Solon201413} discuss potential mechanisms. Some of these
mechanisms require overlapping lifespans or direct contact, such as
when grandparents help in the upbringing of their grandchildren (\citealt{been2022prolonged}),
encourage or pay for educational investments, or transfer wealth.
Other mechanisms do not: grandchildren may still benefit from the
former contacts or reputation of their deceased grandparents, or may
consider them as reference points guiding their own behaviour (\citealp{HertelSamberg2013}). 

More generally, other members of the extended family may affect child
outcomes. For example, \citet{10.1093/esr/jcy021} find that in both
US and Finnish data, aunts and uncles are better predictors of child
education and earnings than grandparents, conditional on parent status.
They note that this observation could be consistent with a model in
which extended family members help with educational investments, provide
access to educational opportunities and jobs, or serve as role models.
And while equation (\ref{eq:1_level_e}) assumes linear effects, a
more realistic model might account for interactions between different
family members (e.g., the extended family might compensate for a lack
of resources in the nuclear family, see \citealt{Jaeger:2012aa} or
\citealp{erola2017book}), the overlap in lifespans, or the size of
the extended family network (\citealp{LehtiErolaTanskanen2018}).

\subsection{Non-linear transmission processes and multiplicity\label{subsec:Non-linear-transmission-processe}}

A third important reason why multigenerational correlations may decay
less rapidly than iterations of the parent-child correlation are non-linearities
and other forms of heterogeneity in the transmission processes. To
see this, consider a transmission process with multiple transmission
pathways (``multiplicity''). Specifically, we introduce a second
endowment into the latent factor model,
\begin{align}
y_{t} & =\rho_{1}e_{1t}+\rho_{2}e_{2t}+u_{t}\label{eq:2_level_2_factors_y}\\
e_{1t} & =\lambda_{1}e_{1,t-1}+v_{1t}\label{eq:2_level_2_factors_e}\\
e_{2t} & =\lambda_{2}e_{2,t-1}+v_{2t},\label{eq:2_level_2_factors_a}
\end{align}
assuming that the two endowments are inherited from parents according
to transferability $\lambda_{1}$ and $\lambda_{2}$. For simplicity,
assume that the noise terms $v_{1t}$ and $v_{2t}$ are uncorrelated,
such that $Cov(e_{1t},e_{2t})=0$ $\forall$ $t$, and that both endowments
affect income ($0<\rho_{1}<1$ and $0<\rho_{2}<1$). The parent-child
and grandparent-grandchild correlations then equal
\begin{align*}
\beta_{-1} & =\rho_{1}^{2}\lambda_{1}+\rho_{2}^{2}\lambda_{2}\\
\beta_{-2} & =\rho_{1}^{2}\lambda_{1}^{2}+\rho_{2}^{2}\lambda_{2}^{2}\\
 & ...
\end{align*}
and so on. In the special case in which incomes are \textit{perfectly}
determined by individual endowments, such that $\rho_{1}^{2}+\rho_{2}^{2}=1$
and $Var(u_{t})=0$, the extrapolation error from the iteration of
the parent-child correlation can be written as (\citealp{Stuhler2012})
\begin{align}
\Delta & =\rho_{1}^{2}(\rho_{1}^{2}-1)(\lambda_{1}-\lambda_{2})^{2},\label{eq:2_levels_2_factors_extrperror-2}
\end{align}
which is negative for $\lambda_{1}\neq\lambda_{2}$. This result
can be understood as the application of Jensen's inequality: the square
of the average transferability across endowments is smaller than the
average of their square. Inequalities therefore decay more slowly
if intergenerational income persistence stems from multiple causal
pathways with differing rates of persistence. 

Figure \ref{Figure_simulations}b provides a numerical example. As
in the latent factor model, multigenerational correlations (solid
blue line) decay more slowly than an iteration of parent-child correlations
(dashed blue line) would suggest. Moreover, the more transferable
trait explains an increasing share of the long-run persistence in
income (dotted line). One interesting implication is that \emph{multi}generational
persistence might reflect factors that are less important in explaining
\emph{inter}generational persistence; in our illustration, endowment
$e_{1}$ contributes less to parent-child transmission than endowment
$e_{2}$, but causes most of the multigenerational persistence. For
example, racial segregation could be an important factor contributing
to multigenerational persistence (\citealt{HertelSamberg2013}, \citealt{Margo2016Journal}).

\begin{figure}[t]
\caption{Multigenerational correlations in different models (continued)}

\label{Figure_simulations2}

\subfloat[A non-linear model with ``poverty traps'']{
\centering{}\includegraphics[scale=0.75]{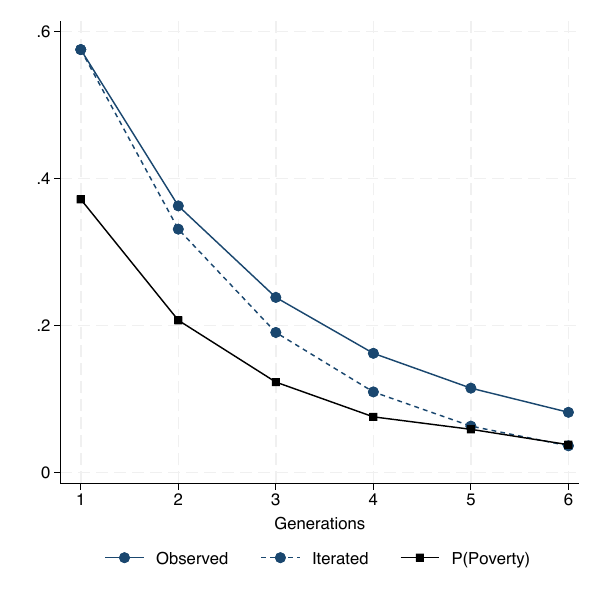}}\subfloat[A latent factor model with assortative mating]{
\centering{}\includegraphics[scale=0.75]{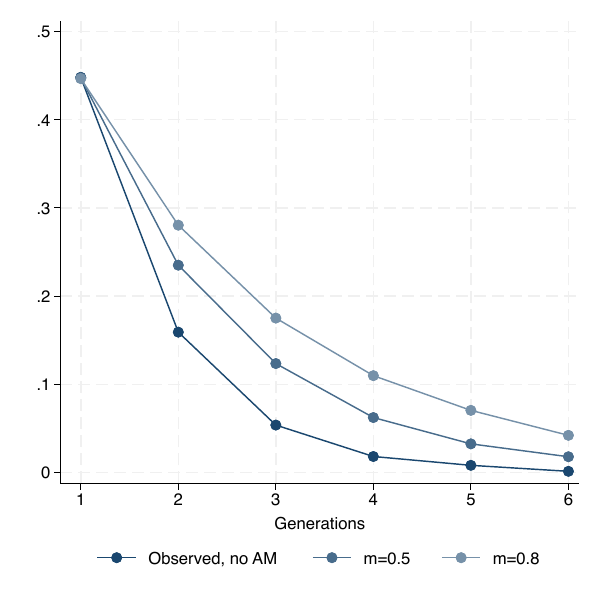}}

{\footnotesize{}\vspace{-0.5cm}
Notes: Simulated data. Panel (a) corresponds to the model in equation
(\ref{eq:poverty traps}) with $\gamma_{1}=0.9$, $\gamma_{2}=0.2$
and $\underline{y}=-0.3$. The solid black line corresponds to the
persistence in poverty ($y<\underline{y}$). Panel (b) reports the
correlations in the observed outcome $y$ in a latent factor model
with assortative mating (see Section \ref{sec:Assortative-Matching}),
assuming $\rho=0.8$, $\widetilde{\lambda}=0.7$ and $m=\{0,0.5,0.8\}$. }{\footnotesize\par}
\end{figure}

Similar implications arise with other forms of heterogeneity, such
as non-linearities in the transmission process, as can be generated
by neighbourhood effects or ``poverty traps'' (see \citealp{durlauf1994spillovers},
and \citealp{nolan2024intergenerational}. For a simple illustration,
assume there is modest transmission in the upper parts of the income
distribution but high persistence in the bottom below some threshold
$\underline{y}$, i.e.
\begin{equation}
y_{t}=\gamma_{1}I(y_{t-1}<\underline{y})\left(y_{t-1}-\underline{y}\right)+\gamma_{2}I(y_{t-1}\geq\underline{y})\left(y_{t-1}-\underline{y}\right)+u_{t}\label{eq:poverty traps}
\end{equation}
where $\gamma_{1}>\gamma_{2}$. Figure \ref{Figure_simulations2}a
provides a numerical example; we again find excess persistence in
the form of $\Delta<0$ in multigenerational correlations (blue solid
line) compared to the prediction based on iterated parent-child correlations
(dashed line). The probability to remain poor, in the sense of $y<\underline{y}$,
decays slowly as well (black line). 

The observation of high multigenerational correlations may therefore
be a consequence of our tendency to ignore non-linearities and heterogeneity
in the parent-child transmission process. But while such non-linearities
have received much consideration in other contexts, its implications
for multigenerational transmission have received less attention. Important
recent exceptions are \citet{BingleyCappellari2019correlation}, who
note that the strength of transmission may vary systematically across
families or different groups, \citet{Colagrossi:2020aa}, who note
that such heterogeneity would affect the relative size of intergenerational
and sibling correlations, and \citet{benhabib2022heterogeneous},
who show that a model with permanent heterogeneity in wealth returns
can match the wealth distribution in both the short- and long-run. 

\subsection{Testing theories of multigenerational transmission\label{subsec:Testing-theories}}

Very different mechanisms could therefore explain similar multigenerational
patterns. Indeed, different fields have emphasised different interpretations.
The latent-factor interpretation has been popular in economics (e.g.,
\citealt{Clark2014book}, \citealp{ECOJ:ECOJ12453}, \citealt{colagrossi2020like}).
The grandparent-effect interpretation has received particular interest
in demography, after \citeauthor{Mare2011} (\citeyear{Mare2011},
\citeyear{Mare2014}) called attention to the role of the wider family.
Both interpretations are found in sociology (\citealp{Tak-Wing-Chan:2013aa};
\citealp{Engzell:2020aa}), while the implications of non-linear transmission
for multigenerational mobility have received little attention in any
field. 

How can we then distinguish between these distinct interpretations?
First, the observation of multigenerational persistence as such does
not point to any particular theory. In particular, the finding that
$\beta_{gp}>0$ when estimating equation (\ref{eq:3-generation-reg})
should not be interpreted as favouring the grandparent-effect interpretation.
As follows from (\ref{eq:Duality}), \emph{any} process that generates
high multigenerational persistence in the sense of $\beta_{-2}>(\beta_{-1})^{2}$
will also generate a positive grandparent coefficient, and vice versa.
See also \citet{Lundberg2020}, who makes a similar point regarding
the observation that cousin correlations tend to be larger than the
square of the sibling correlation.

Researchers therefore need more specific evidence to distinguish between
different candidate models. One testable implication of the latent
factor model is that the coefficient $\beta_{gp}$ should be sensitive
to which parental characteristics are controlled for.\footnote{See also a recent strand of the literature that combines multiple
proxy measures of parental status to explain child outcomes (\citealt{VostersNybom2017},
\citealt{BlundellRisa2018aa}, \citealt{hsu2021intergenerational}
or \citealt{eshaghniawellbeing}), or the literature on inequality
of opportunity that often considers a wide set of ``circumstances''
(\citealt{brunori2013inequality}, \citealt{BrunoriHufeMahlerRoots}).} \citet{WarrenHauser1997} show that after conditioning on multiple
parental characteristics, the occupational status of grandparents
is not predictive of child status. Many recent studies report similar
``conditioning tests'' (see Table \ref{table:studies}), typically
finding that $\beta_{gp}$ shrinks but remains positive after controlling
for a wide set of parental characteristics (\citealp{SheppardMonden2018}).
Such residual associations are still consistent with a latent factor
model, as relevant characteristics might be missing in the data at
hand, or be fundamentally unobservable. 

Indeed, the more relevant question may be by \emph{how much} $\beta_{gp}$
declines when controlling better for parental characteristics, rather
than whether $\beta_{gp}$ remains positive. Using rich data from
Sweden, \citet{Engzell:2020aa} show that even models that control
for both parents\textquoteright{} education, earnings, occupation,
and wealth may still suffer from bias from the omission of relevant
parental characteristics.\footnote{Related, \citet{modalsli2022spillover} show that measurement in parent
income may generate a spurious grandparent coefficient, even if a
long-term average of parental income is controlled for.} Moreover, while conditioning tests suggest that ``grandparent effects''
largely reflect omitted parental variables, \citet{Breen2018} notes
that these tests are subject to interpretative issues, too.\footnote{\citet{Breen2018} shows that \emph{partial} conditioning for some
but not all relevant characteristics of the parent generation may
not always reduce bias. Using the language of causal graphs, the issue
is that parental status might not only be a ``mediator'' for the
effect of grandparent on child status, but also a ``collider'' that
is affected by other causal factors influencing both parent and child
status (such as neighbourhood effects).}

One obvious source of omitted variable bias is the omission of one
of the parents: the coefficient $\beta_{gp}$ on grandparent\textquoteright s
status tends to be smaller when controlling for \emph{both} paternal
and maternal characteristics. \citet{ECOJ:ECOJ12453} show that in
German samples, the coefficient on grandfather's status declines strongly
once we condition on the corresponding status of the mother; indeed,
\citet{NeidhoeferStockhausen} and \citet{Engzell:2020aa} find that
the coefficient can become negligible once both parents are accounted
for. Detailed conditioning tests are also provided by \citet{chiang2015grandparents},
\citet{Fiel2018} or \citet{SheppardMonden2018}.

Another interesting test is whether the association between grandparent
and child status varies with the extent of contact between them. In
a systematic review of the literature, \citet{AndersonSheppardMonden2017}
find that the coefficient $\beta_{gp}$ does not appear to vary systematically
with the likelihood of contact between grandparent and grandchild
(see also \citealp{HelgertzDribe2021}). However, \citet{ZengXie2014},
\citet{Knigge2016} and \citet{song2019shared} are notable exceptions,
and a direct effect of grandparents on their grandchildren is also
possible through means that do not require contact, such as financial
transfers (\citealp{hallsten2017grand}). 

A third strategy is to estimate persistence at an aggregate rather
than individual level. For example, the parameter $\lambda$ in the
latent factor model could be estimated by averaging $y$ across multiple
relatives, to then estimate the regression to the mean on the surname
level. With this motivation in mind, \citet{Clark2014book}, \citet{ClarkCummins2014EJ}
and related studies document strikingly strong persistence of socio-economic
status at the \emph{surname} level, across many different countries
and time periods. While the precise interpretation of name-based estimators
is contested (e.g. \citealt{RocheCorvalan2016}, \citealt{SantavirtaStuhler2019aa}),
they might prove useful to discriminate between competing models of
multigenerational inequality. For example, \citet{Belloc2023uy} find
that even great-grandparents' wealth is still predictive of child
wealth, conditional on parents' and grandparents' wealth, and note
that this pattern could be consistent with a latent factor model. 

A fourth strategy is to confront competing transmission models with
a wider set of kinship correlations. The ``right'' model should
explain not only the relation between multi- and intergenerational
correlations, but also their relation to sibling and many other type
of kinship correlations. For example, \citet{ColladoOrtunoStuhlerKinship},
show that a generalised latent factor model can provide a good fit
to a wide set of 141 distinct kinship moments in Swedish data. One
interesting question is whether non-linear or non-Markovian models
could provide a similarly good fit, or explain status correlations
between very distant ancestors (\citealp{BaroneMocetti2016}).

These obstacles in the interpretation of multigenerational correlations
are of course the same obstacles that limit our understanding of social
mobility more generally. Statistical associations are difficult to
map to mechanisms, and while robust causal evidence exists for certain
pathways, most of the statistical associations remain unaccounted
for (\citealp{bjorklund2020intergenerational}). Moreover, distinguishing
between the models considered in Section \ref{sec:Interpretations}
is only a first challenge, as those stylised models are not precise
about the \emph{specific} mechanisms and behavioural patterns that
matter, or how policies and institutions would affect them. 

Still, the recent multigenerational evidence is not ``toothless'',
as it reduces the range of permissible models. For example, the standard
implication of the Becker-Tomes model for $\beta_{gp}$ to be negative
(see Section \ref{subsec:Latent-transmission-processes}) is rejected
by most papers. Related, \citet{ColladoOrtunoStuhlerKinship} show
that a purely genetic model with phenotypic assortment could not fit
the kinship pattern in educational advantages. And while there is
no consensus yet on the underlying mechanisms, multigenerational estimates
are directly informative about the extent of status persistence in
the long run -- thereby providing novel insights about an important
dimension of inequality. 

\subsection{Does this matter? The $R^{2}$ controversy\label{sec:Caveats-and-Criticisms}}

Do multigenerational associations \textquotedblleft matter\textquotedblright{}
in a quantitative sense? One frequent observation is that conditional
on parents, other ancestors do not add much to the regression $R^{2}$
-- even if the corresponding slope coefficients appear sizeable.
For example, \citet{PfefferKillewald2018} report that switching from
a two- to a three-generations regression, the $R^{2}$ increases only
mildly (from 0.146 and 0.160), even though the coefficient on the
grandparents' wealth is nearly half as large as the coefficient on
the parents' wealth. \citet{ErolaMoisio2007} make a similar observation
in Finnish data. This low contribution in a $R^{2}$ sense is also
a key point in an interesting recent debate between \citet{bjorklund2022comment}
and \citet{adermon2022comment}.

To provide an illustration, we generate simulated data based on the
latent factor model given by (\ref{eq:1_level_y}) and (\ref{eq:1_level_e})
and the same parameters as in Figure \ref{Figure_simulations} ($\lambda=0.7$
and $\rho=0.8$). Table \ref{tab:r2} reports regression estimates
from this simulated data. Column (1) reports estimates of the intergenerational
coefficient $\beta_{-1}$, which according to equation (\ref{eq:2_layers_betaIGE})
equals $\beta_{-1}=\rho^{2}\lambda$. In column (2) we add the grandparent
status to the model. But while the coefficient on grandparents is
sizeable, the regression $R^{2}$ hardly increases. Hence the conundrum:
are deviations from the ``iterated'' parent-child regression (Section
\ref{sec:The-iterated-regression}) important, as indicated by the
regression slopes, or negligible, as seemingly implied by the $R^{2}$? 

\begin{table}
\caption{On the explanatory power of multigenerational associations}

\begin{centering}
\label{tab:r2}%
\begin{tabular*}{1\textwidth}{@{\extracolsep{\fill}}lcccccc}
\toprule 
 & \multicolumn{6}{c}{Dependent variable: Child status $y$}\tabularnewline
 & (1) & (2) & (3) & (4) & (5) & (6)\tabularnewline
\midrule
Parent's $y$ & 0.450{*}{*}{*} & 0.389{*}{*}{*} & -- & 0.392{*}{*}{*} & -- & 0.310{*}{*}{*}\tabularnewline
 & (0.004) & (0.004) &  & (0.004) &  & (0.004)\tabularnewline
Grandparent's $y$ & -- & 0.138{*}{*}{*} & -- & -- & -- & --\tabularnewline
 &  & (0.004) &  &  &  & \tabularnewline
Sibling's $y$ & -- & -- & 0.306{*}{*}{*} & 0.131{*}{*}{*} & 0.458{*}{*}{*} & 0.318{*}{*}{*}\tabularnewline
 &  &  & (0.004) & (0.004) & (0.004) & (0.004)\tabularnewline
$R^{2}$ & 0.204 & 0.219 & 0.095 & 0.218 & 0.210 & 0.286\tabularnewline
\bottomrule
\end{tabular*}
\par\end{centering}
{\footnotesize{}\smallskip{}
Notes: Simulated data from the latent factor model in equations (\ref{eq:1_level_y})
and (\ref{eq:1_level_e}) with returns $\rho=0.8$ and transferability
$\lambda=0.7$ over three generations (n=50,000). For columns (3)
and (4), the noise term $u$ is uncorrelated between siblings. For
columns (5) and (6), $u$ is drawn from a joint normal distribution
with correlation 0.4 between siblings.}{\footnotesize\par}
\end{table}

The answer depends on \emph{why} multigenerational associations exist.
While the $R^{2}$ differs little between columns (1) and (2), the
significant coefficient on grandparents signals that the transmission
process deviates from the simple parent-child regression. And in our
chosen example, that deviation turns out to be important: advantages
are transmitted at a \emph{much} higher rate than the parent-child
correlation suggests ($\lambda=0.7$ vs. $\beta\approx0.45$), and
the multigenerational implications differ substantially (Figure \ref{Figure_simulations}a).
The observation of independent multigenerational associations can
therefore be meaningful, even if the regression $R^{2}$ moves little. 

It is instructive to compare multigenerational to \emph{sibling correlations},
as an alternative measure of family influences. We simulate two children
per parent, initially assuming that the errors $u$ and $v$ in equations
(\ref{eq:1_level_y}) and (\ref{eq:1_level_e}) are uncorrelated between
siblings. As shown in column (3) of Table \ref{tab:r2}, the implied
sibling correlation (0.31) is much larger than the square of the parent-child
correlation ($0.45^{2}=0.20$), even though the addition of siblings
does not add much to the $R^{2}$ in a parent-child regression (cf.
columns 1 and 4). The ``excess persistence'' found in multigenerational
studies (column 2) and sibling correlations (cf. columns 1 and 3)
could therefore reflect similar mechanisms.

However, the measures are not interchangeable: sibling correlations
are a broad measure of family background that also capture environmental
influences that siblings share, such as neighbourhood or peer effects
(\citealp{BjorklundSalvanes2011HB}, \citealt{JenkinsJaenttHandbook201x}),
while multigenerational correlations capture the effect of ancestry
in a narrow sense. To illustrate this point, columns (5) and (6) replicate
the regressions from columns (3) and (4) but allow for shared influences
between siblings (see table notes). This increases the sibling correlation
further (cf. columns 3 and 5), and the $R^{2}$ in a regression with
parents \emph{and} siblings is now substantially larger as compared
to the simple parent-child regression (cf. columns 1 and 6), as is
the case in actual applications.

\section{Assortative Matching\label{sec:Assortative-Matching}}

The observation of significant multigenerational correlations also
has implications for the \emph{assortative} matching between spouses.
To see this, extend the latent factor model as given by equations
(\ref{eq:1_level_y}) and (\ref{eq:1_level_e}) to a two-parent setting
(see also \citealp{ECOJ:ECOJ12453}), assuming that an offspring\textquoteright s
endowment depends on the average of the father\textquoteright s and
mother\textquoteright s endowment,
\begin{equation}
\ensuremath{e_{t}=\tilde{\lambda}\frac{e_{t-1}^{m}+e_{t-1}^{p}}{2}+v_{t}},\label{eq:AM_e}
\end{equation}
where the $m$ and $p$ superscripts denote maternal and paternal
variables, respectively. Normalising $y$ and $e$ to one, the parent-child
correlation in $y$ between a child and \emph{one} parent is now given
by
\begin{equation}
\beta_{-1}=\rho^{2}\tilde{\lambda}\ensuremath{\frac{1+m}{2}},\label{eq:AM_beta1}
\end{equation}
where $m=Corr\left(e_{i,t-1}^{m},e_{i,t-1}^{p}\right)$ measures the
extent of assortative matching among parents. The multigenerational
correlations with earlier ancestors are similarly given by 
\begin{equation}
\beta_{-k}=\rho^{2}\tilde{\lambda}^{k}\left(\ensuremath{\frac{1+m}{2}}\right)^{k}\label{eq:AM_Betak}
\end{equation}
for $k>1$. Because $\beta_{-k}$ depends on $\left(\ensuremath{\frac{1+m}{2}}\right)^{k}$,
multigenerational correlations will decay quickly if assortative matching
is weak -- even if parental endowments were transmitted perfectly.\footnote{Here we consider the correlations with a single ancestor. Some studies
average over all ancestors in a generation, thereby abstracting from
sorting. For example, in the model considered here the correlation
between the child and parents' average status would be $\rho^{2}\tilde{\lambda}$
and thus not depend on $m$. }

Figure \ref{Figure_simulations2}b provides a numerical example with
$\rho=0.8$ and $\tilde{\lambda}=0.7$ and three different assortative
correlations, with $m=0$ (no assortative matching), $m=0.5$ (a typical
value for the spousal years of schooling, \citealp{fernandez2005love})
and $m=0.8$. The example illustrates that high multigenerational
correlations are possible only if assortative matching is very strong,
a point that is also discussed by \citet{Clark2014book}. Indeed,
conventional assortative measures may understate sorting for the same
reasons that parent-child correlations understate intergenerational
transmission: the similarity of spouses in observable characteristics,
such as years of schooling, may not capture their resemblance in unobserved
determinants of child outcomes. \citet{ColladoOrtunoStuhlerKinship}
estimate that in Sweden, the spousal correlation in such latent determinants
must be around 0.75, or more than 50 percent larger than the spousal
correlation in years of schooling, to explain the correlation patterns
between distant in-laws. 

\section{Conclusions}

This chapter provided an overview of the fast-growing literature on
multigenerational inequality. Using data across three or more generations,
recent studies show that socio-economic inequality tends to be more
persistent than a naive extrapolation from conventional parent-child
measures would suggest. However, very distinct interpretations of
that new ``fact'' are possible. Its significance might lie less
in the magnitude of multigenerational associations per se, as adding
earlier ancestors often contributes little to the overall explanatory
power of parent-child models. More importantly, multigenerational
associations tell us something novel about the nature of intergenerational
processes. 

One plausible interpretation is that parent-child transmission --
and assortative mating -- must be much stronger than what conventional
measures capture. But the chapter also reviews alternative interpretations,
related to ``grandparent effects'' or non-linearities in the transmission
process, and more work is needed to distinguish between competing
models. One exciting aspect is that this work is happening simultaneously
in multiple fields of the social sciences, connecting different strands
of research that otherwise tend to progress in isolation. 

To retain focus, we passed over many important aspects. Empirically,
one major concern is sample selection regarding the coverage of different
generations, the coverage of migrants, or the age at which the outcome
of interest can be observed.  Conceptually, some studies rely on
steady-state assumptions that are less plausible when comparing many
generations (\citealp{NybomStuhler2019}), and most abstract from
demographic processes and differences in fertility (\citealp{song2021multigenerational}).
 A more explicit consideration of these aspects could be fruitful
for future work. 

\newpage{}

\begin{spacing}{1.08}
\noindent \pagestyle{empty}\setlength{\bibsep}{0pt}\bibliographystyle{aea}
\bibliography{library_main_stuhler}
\end{spacing}

\end{document}